\documentclass[twocolumn,superscriptaddress,floatfix,
longbibliography,aps,pra,preprintnumbers]{revtex4-1}

\usepackage[utf8]{inputenc}
\usepackage[T1]{polski}
\usepackage[english]{babel}

\usepackage{graphicx}
\usepackage{bm}
\usepackage{xcolor}
\usepackage{epstopdf}
\usepackage{amsmath}
\usepackage{amssymb}
\usepackage{epstopdf}
\usepackage{ulem} 
\usepackage{enumerate}

\usepackage[urlcolor=blue,colorlinks=true,citecolor=blue,linkcolor=blue,pdfstartview={FitH},bookmarks=false]{hyperref}

\definecolor{darkgreen}{HTML}{008000}

\graphicspath{{fig/}{./fig/}{.}}

\begin{document}

\title{Majorana bound states and zero-bias conductance peaks in superconductor/semiconductor nanowire devices}

\author{Aksel Kobia\l{}ka}
\email[e-mail: ]{akob@kft.umcs.lublin.pl}
\affiliation{Institute of Physics, Maria Curie-Sk\l{}odowska University, 
Plac Marii Sk\l{}odowskiej-Curie 1, PL-20031 Lublin, Poland}

\author{Andrzej Ptok}
\email[e-mail: ]{aptok@mmj.pl}
\affiliation{Institute of Nuclear Physics, Polish Academy of Sciences, 
ul. W. E. Radzikowskiego 152, PL-31342 Krak\'{o}w, Poland}

\date{\today}

\begin{abstract}
Theoretical research suggests a emergence of the Majorana bound states at the ends of the nanowires. 
Experimental verifications of said concept has already been executed, e.g., in superconductor/semiconductor nanowire devices where interplay between superconducting gap, spin--orbit coupling and external magnetic field allows for creation of zero--energy bound states.
Recent experiments propose a topological phase diagram by local modification of the effective chemical potential.
We discuss this possibility, using a model of experimental system in form of semi--infinite S/N junction.
We calculate the zero--bias differential conductance $G$ in the case of the homogeneous system, as well as in the presence of the gate voltage.
Relation between conductance and the effective gap in the system is investigated. 
We show that $G$ can reproduce the topological phase diagram in magnetic field vs. gate voltage space of parameters.
\end{abstract}


\maketitle

\section{Introduction}

Proposed by Kitaev in 2001, a creation of the Majorana bound states (MBS)~\cite{kitaev.01} is very attractive concept, due to its non-Abelian properties~\cite{nayak.simon.08}, which can be using in performing of the quantum computation via braiding operation~\cite{alicea.oreg.11,aasen.hell.16,wieckowski.mierzejewski.19}.
This proposal opened a period of intensively theoretical and experimental studies of the MBS~\cite{aguado.17,lutchyn.bakkers.18,pawlak.hoffman.19}.

In Kitaev toy model~\cite{kitaev.01}, the MBS emerge at the end of the chain of spinless fermions with inter--site paring.
From theoretical point of view, in the superconducting system with the spin--orbit coupling, a mixture of both spin--singlet and spin--triplet Cooper pairs can be expected~\cite{gorkov.rashba.01,seo.han.12,ptok.rodriguez.18}.
In the presence of the significantly high magnetic field, the pairing in only one branch exists~\cite{potter.lee.11}, which is formally equivalent to a spinless p$_{x}$+ip$_{y}$ superconductor~\cite{fu.kane.08}.

Experimentally, this situation can be performed in the superconductor/semiconductor heterostructures~\cite{lutchyn.bakkers.18}.
In those types of systems, interplay between intrinsic spin--orbit coupling of semiconductor, superconducting gap induced in semiconductor (by the proximity effect with superconductors~\cite{chang.albrecht.15,gul.zhang.17}) and external magnetic field can lead to emergence of MBS~\cite{mourik.zuo.12}.

\begin{figure}[!b]
\centering
\includegraphics[width=\linewidth]{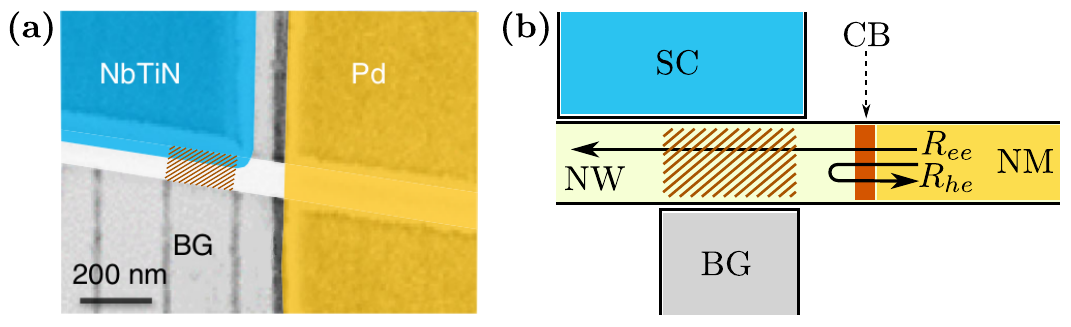}
\caption{
(a) Scanning electron micrograph of the superconductor/semiconductor nanowire devices studied in Ref.~\cite{chen.woods.19}.
(b) Theoretical model of the experimental system:
semiconducting nanowire (NW) coupled to the normal metal (NM) lead and at the junction  Coulomb blockade (CB) exists, NW is coupled to the superconductor (SC), energy levels are shifted by the background gate (BG) voltage.
\label{fig.schemat}
}
\end{figure}

Example of practical assembly of superconductor/semiconductor nanowire devices is presented at Fig.~\ref{fig.schemat}(a). 
Semiconducting nanowire (white) is partially covered by superconducting (blue) and metallic lead (orange).
This is the procedure generally used in obtaining the heterostructures to realization of the MBS~\cite{mourik.zuo.12,deng.vaitiekenas.16,zhang.gul.17,chen.yu.17,chen.woods.19}.
Moreover, recent progress in experimental techniques in preparation of heterostructures, allow for construction additional gates~\cite{das.ronen.12,deng.vaitiekenas.18}.
These gates [like background gate (BG) at Fig.~\ref{fig.schemat}] can be used to locally modify of the effective chemical potential via electrostatic--control.
Experimentally, this can be used to reproduce the topological phase diagram~\cite{chen.yu.17}.

In the superconducting system with spin--orbit coupling, increase of magnetic field leads to the closure of gap and reopening new topological gap.
This occurs in some magnetic field $h_{c} = \sqrt{ \mu^{2} + \Delta^{2} }$~\cite{sato.fujimoto.09,sato.takahashi.09,sato.takahashi.10}, where $\mu$ is the chemical potential (measured form bottom of the band), while $\Delta$ is a superconducting gap in the absence of the magnetic field.
In practice, chemical potential is fixed for given materials, while electron concentration can be changed with gate potential~\cite{chen.yu.17}.
In this paper, we examine this concept using the effective Bogoliubov--de~Gennes tight-binding model of the one dimensional (1D) model of experimental system.
The paper is organized as follows.  
In Sec.~\ref{sec.model} we introduce the microscopic lattice model and present technique.
In Sec.~\ref{sec.num_res}, we describe the numerical results.
Finally, we summarize the results in Sec.~\ref{sec.sum}.

\section{Model and technique}
\label{sec.model}

We focus on the single-band one-dimensional approximation, which is valid when the electron occupation is low and the interband spacing is large, compared to other energy scales in the problem~\cite{stanescu.tewari.19}.
The low-energy physics of the hybrid devices can be relatively well captured by this approximation~\cite{liu.sau.17,huang.pan.18,pan.dassarma.19,stanescu.tewari.19}.
We will describe experimental system [Fig.~\ref{fig.schemat}(a)] using a semi-infinite model of the superconducting/normal (SN) junction, which is schematically shown at Fig.~\ref{fig.schemat}(b).
Here, the semiconducting nanowire (NW) is connected with normal metal (NM) lead. 
At the connection between NW and NM Coulomb blockade (CB) region forms~\cite{aleiner.brouwer.02}.
Coupling of the NW with the superconductor (SC) leads to the proximity induced superconducting gap  $\Delta$ (in calculations we assume constant value of gap).
Additionally, in some part of NW we can modify local chemical potential using background gate (BG) voltage $V_\text{BG}$ (brown region in Fig.~\ref{fig.schemat}).

In the absence of the gate, described system can be emulated by the effective Bogoliubov--de~Gennes tight-binding model~\cite{ptok.kobialka.17}:
\begin{eqnarray}
\nonumber H_{0} &=& \sum_{i,j,\sigma} \left[ - t \delta_{\langle i,j\rangle} - ( \tilde{\mu} + \sigma h ) \delta_{ij} \right] c_{i\sigma}^{\dagger} c_{j\sigma} \\
\label{eq.ham_re} &+& i \lambda \sum_{i\sigma\sigma'} \left( c_{i+{\bm x},\sigma}^{\dagger} \sigma_{y}^{\sigma\sigma'} c_{i\sigma'} + h.c. \right) \\
\nonumber &+& \Delta \sum_{i} \left( c_{i\uparrow}^{\dagger} c_{i\downarrow}^{\dagger} + h.c. \right) ,
\end{eqnarray}
where $\langle i,j\rangle$ are the nearest-neighbor sites in the lattice, $c_{i\sigma}^{\dagger}$ ($c_{i\sigma}$) is the electron creation (annihilation) operator of electron with spin $\sigma$ in site $i$.
Here, $t$ is the hopping integral, $\tilde{\mu} = \mu - 2t$ is the on-site chemical potential (measured from the bottom of the band), $h$ is the Zeeman magentic field, $\lambda$ is the Rashba spin-orbit coupling, while $\sigma_{y}$ is a Pauli matrix.

We assume that the gate potential acts on the electron locally, only in some part of NW around the BG.
Thus, the gate potential $V_\text{BG}$ is given in the on-site term form:
\begin{eqnarray}
H_\text{gate} =  \sum_{i} V_\text{BG} ( i ) c_{i\sigma}^{\dagger} c_{i\sigma} ,
\end{eqnarray}
where we assume $V_\text{BG} ( i ) \equiv V_\text{BG}$ for sites in the BG region, and $V_\text{BG} ( i ) \equiv 0$ otherwise.
From this, chemical potential is locally modified by $V_\text{BG}$.
Therefore, effective chemical potential has form $\tilde{\mu}_{i} \rightarrow \tilde{\mu} - V_\text{BG} ( i )$.


We will calculate the conductance $G$ of system using $S$ matrix method~\cite{lesovik.sadovskyy.11,akhmerov.dahlhaus.11,fulga.hassler.11}.
In particular, the conductance of a S/N junction is given as:
\begin{eqnarray}
G = \frac{e^{2}}{h} \left( N - R_{ee} + R_{he} \right) ,
\end{eqnarray}
where $N$ is the number of electron channels in the normal lead, $R_{eh}$ is the total probability of reflection from electron to holes in the normal lead, while $R_{ee}$ is the total probability of reflection from electrons to electrons in the normal lead.
We performed the calculations in the semi-infinity S/N junction~\cite{liu.sau.17,huang.pan.18}, using the {\sc Kwant} code~\cite{groth.wimmer.14}.
In typical situation $G$ is quantized by $G_{0} = e^{2}/h$~\cite{wimmer.akhmerov.11}.
However, in the case of the ``true'' MBS measured of zero-energy bias peak with $G = 2 G_{0}$ is expected~\cite{wimmer.akhmerov.11,kjaergaard.nichele.16}.
This should be treated as a signature of realization of the MBS in the system~\cite{zhang.liu.18}.

\section{Numerical results}
\label{sec.num_res}

In our calculations we take $\Delta/t = 0.3$ and $\tilde{\mu} = 0$ in the case of the homogeneous nanowire $V_\text{BG}/t = 0$.
Calculation was performed for nanowire with length $L=100$ and $200$, what corresponds to the BG region for $i \in ( 50; 80 )$ and $(100; 180)$, respectively.

\begin{figure}[!b]
\centering
\includegraphics[width=\linewidth]{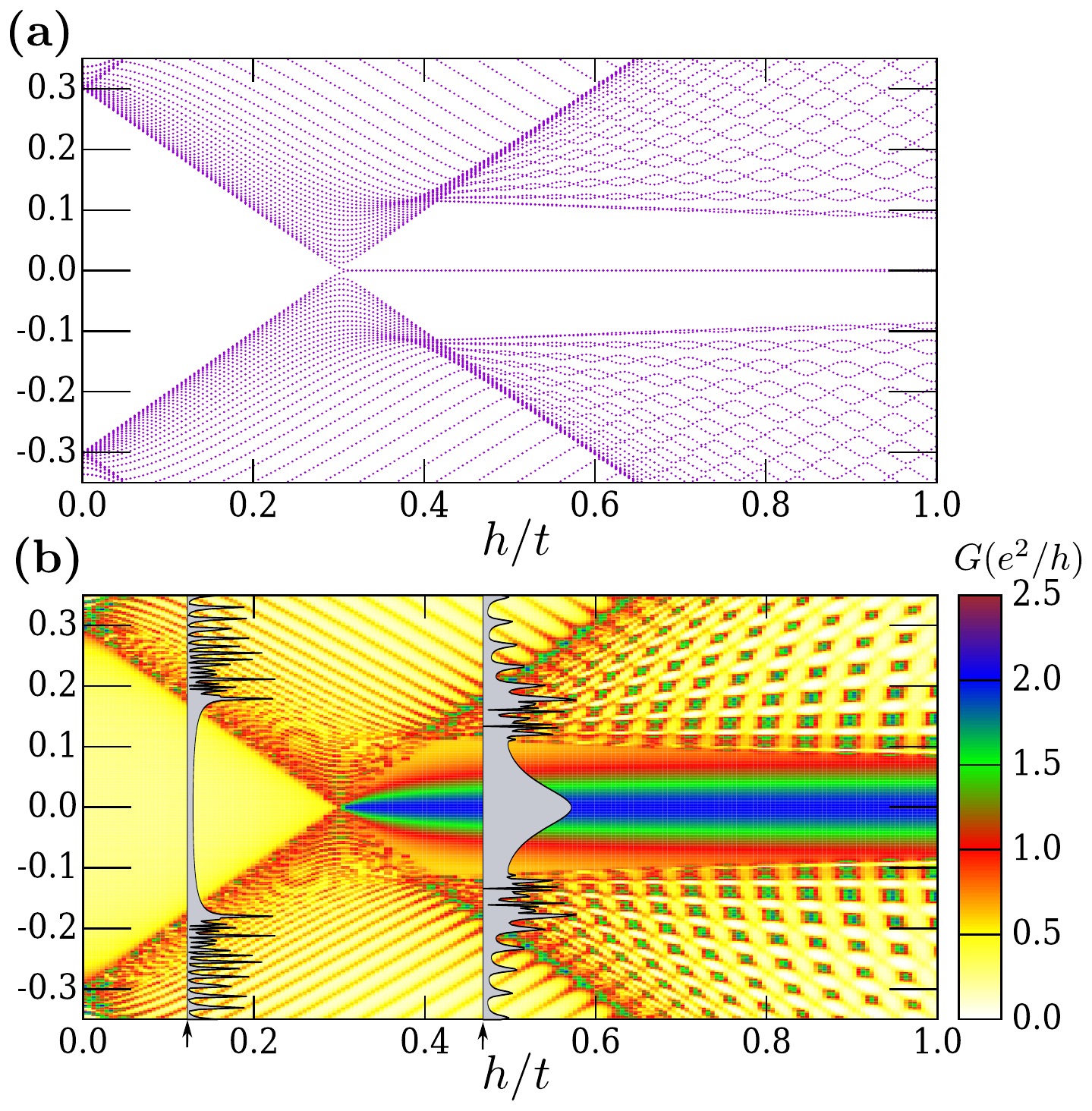}
\caption{
(a) Spectrum of the system in function of the magnetic field $h$.
(b) The differential conductance $G$ in function of the bias voltage and the magnetic field $h$.
For two values of $h$ we have shown contour of $G$ (black lines).
Results for the homogeneous system ($\mu = 0$) with $L = 100$.
\label{fig.res3}
}
\end{figure}

First, we discuss the general properties of the system (Fig.~\ref{fig.res3}).
Spectrum of the system is shown at Fig.~\ref{fig.res3}(a) and represents characteristic features of investigated system~\cite{sticlet.bena.12,kiczek.ptok.17,kobialka.ptok.19b}.
We observe topological phase transition in some magnetic field (for $\mu = 0$ transition occur in $h_{c} \simeq \Delta$).
At this magnetic field ($h = h_{c}$), band inversion occurs~\cite{kobialka.ptok.19} what is an evidence of the transition from trivial to non-trivial phase.
In the case of the system with edge, inside topological gap (for $h > h_{c}$) we observe existence of one pair of the MBS with significantly small (close to zero) eigenenergies .
States outside of the topological gap show oscillating character that leads to its crossing~\cite{sticlet.bena.12,kiczek.ptok.17,kobialka.ptok.19b}.
Spectrum of the system determined from of the differential conductance $G$ [Fig.~\ref{fig.res3}(b)].
Similarly to previous study, we observe increase of $G$ intensity in the for the crossing of  eigenvalues~\cite{rainis.trifunovic.13}.

\begin{figure}[!t]
\centering
\includegraphics[width=\linewidth]{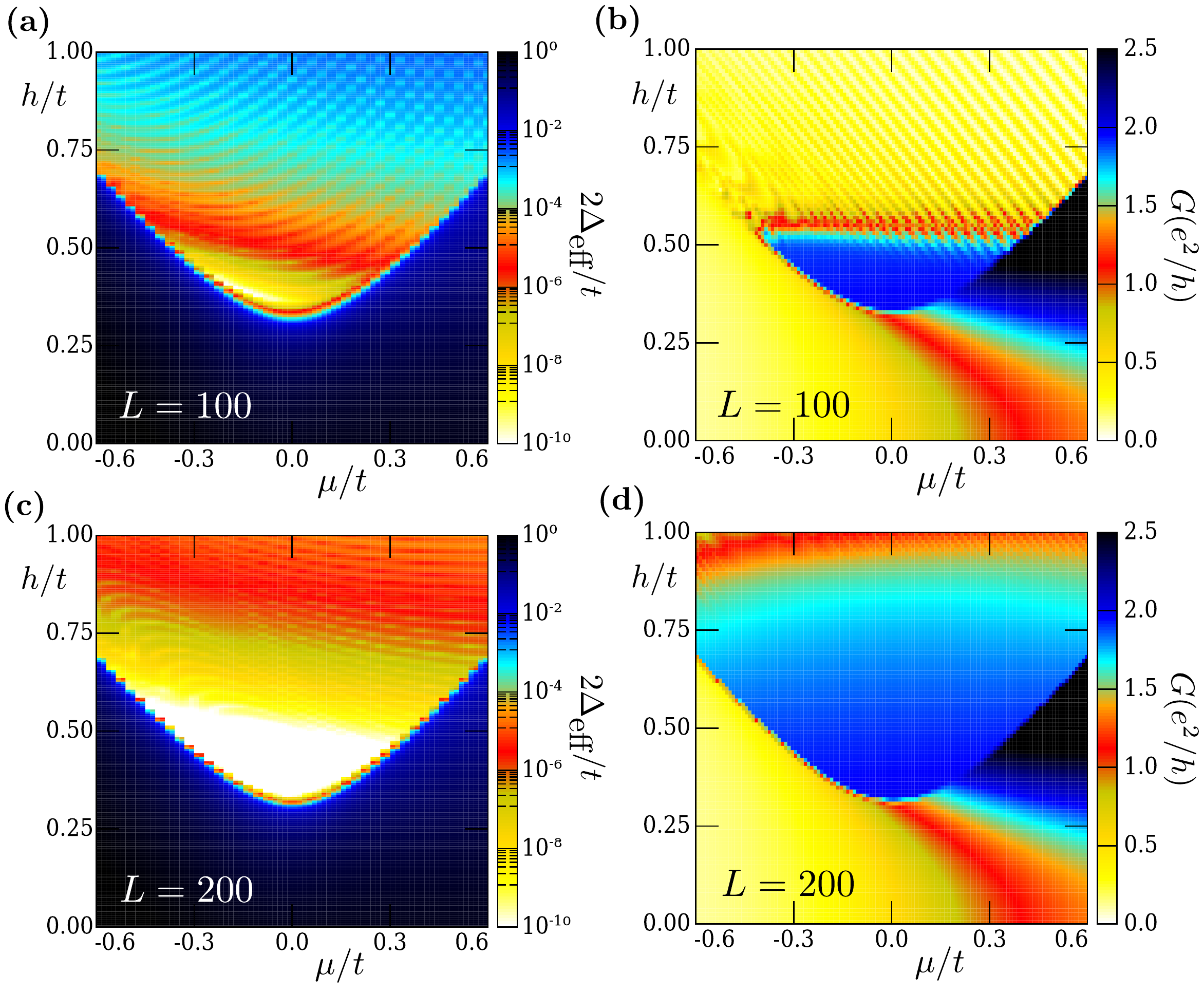}
\caption{
(a) Effective gap values $2\Delta_\text{eff}$ given as a difference between two eigenvalues nearest zero energy.
(b) Zero-bias differential conductance $G$.
Results for the homogeneous system.
\label{fig.res}
}
\end{figure}

At Fig.~\ref{fig.res3} we shown also two cross--sections of $G$ for two values $h$ below and above $h_{c}$.
For $h < h_{c}$ we observe a relatively small value of $G$ inside trivial gap.
Situation looks different in the case of $h > h_{c}$, where inside topological gap, we observe clearly visible increase of $G$ associated with existence of the MBS states.

Now we discuss results in the case of the homogeneous system (Fig.~\ref{fig.res}).
At left panels we show values of effective gap $2\Delta_\text{eff}$, taken as a difference between two eigenvalues of the system, nearest to the Fermi level.
For fixed chemical potential $\tilde{\mu}$ increasing the magnetic field to $h_{c}$ leads to the closing  of trivial gap.
At the $h = h_{c}$, the topological phase transition occurs and new topological gap is reopened.
However, in the finite system, for $h > h_{c}$ the exponential suppression of the gap value is observed.
It is a consequence of the existence of MBS localized at the end of the chain, with energy $\propto \exp ( -L/\xi )$~\cite{sarma.freedman.15}, where $L$ is the system size, while $\xi$ is the correlation length.
Additionally, for larger values of $h$ typical oscillation of the $2\Delta_\text{eff}$ is observed, due to oscillating dependence of the energy levels around the Fermi level~\cite{dassarma.sau.12,escribano.yeyati.18}.
Magnetic field dependence of $2\Delta_\text{eff}$ have strong influence on the zero-bias conductance $G$, which is shown at the right panels.
Independently to the length $L$ of the system, in the topological phase ($h > h_{c}$) we observe $G = 2 G_{0}$, which is an evidence of existence of the true MBS in the system. 
This value exists for the broad range of parameters $h$--$\mu$, where the $2\Delta_\text{eff}$ is significantly small, i.e. typically when $2\Delta_\text{eff}/t$ is smaller than $\sim 10^{-7}$, what corresponds to the magnetic field $h/t < 0.55$ and $0.75$ in the case of $L = 100$ and $200$, respectively.
Independently to this, in both cases, a boundary between trivial and non-trivial part of the phase diagram is clearly visible and takes the form of parabolas $h_{c} = \sqrt{ \mu^{2} + \Delta^{2} }$.

\begin{figure}[!t]
\centering
\includegraphics[width=\linewidth]{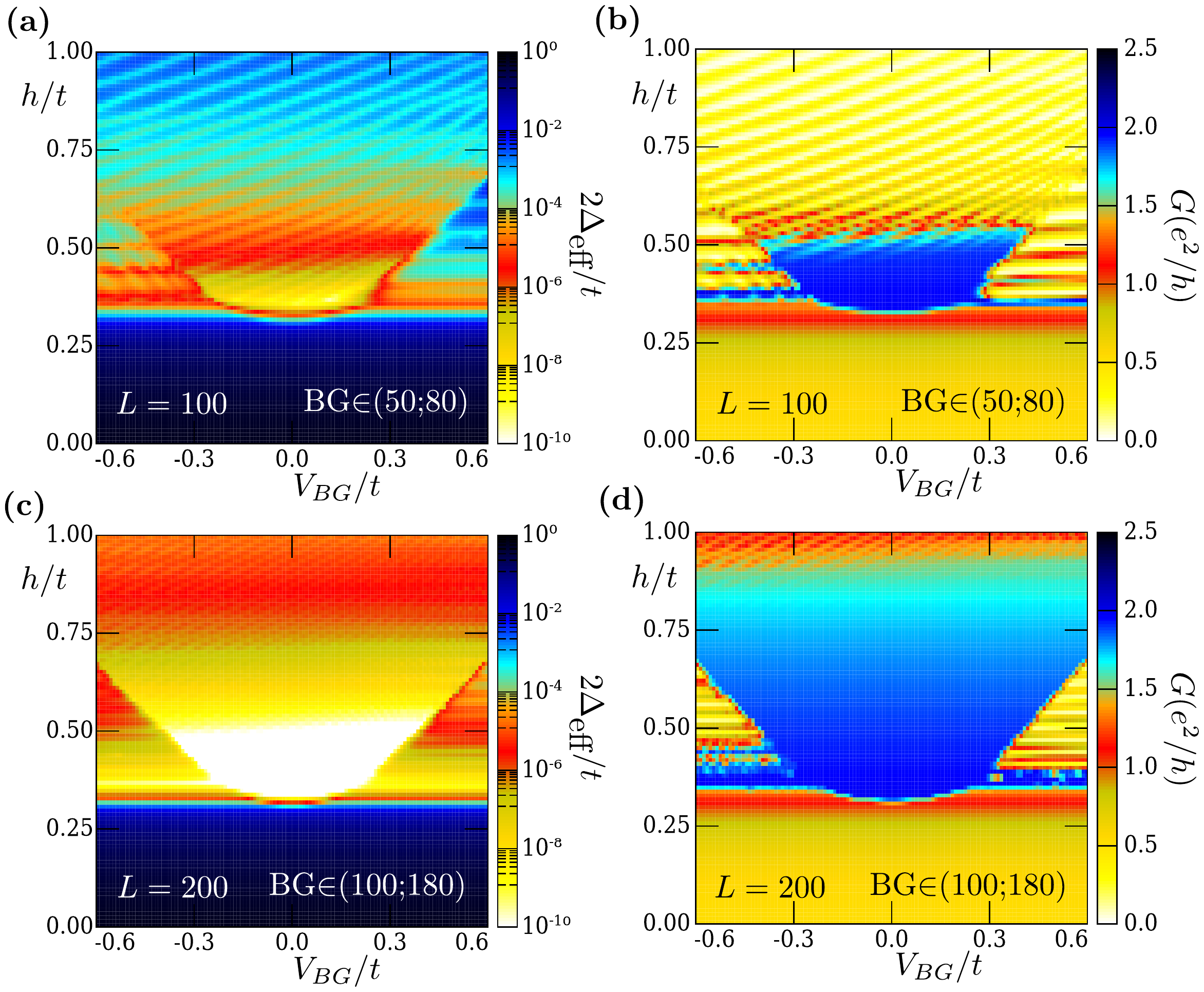}
\caption{
(a,c) Effective gap values $2\Delta_\text{eff}$ given as a difference between two eigenvalues nearest zero energy.
(b,d) Zero--bias differential conductance $G$.
Results for the non-homogeneous system.
\label{fig.res2}
}
\end{figure}

Now, we will discuss a non-homogeneous system -- in which a part of nanowire has its effective chemical potential modified by the BG voltage.
We also probe the system of different length of the BG region.
For $L = 100$ and $200$ we take BG region for $i \in ( 50;80)$ and $(100;180)$, respectively.
Here, it should be mentioned, that the length of the BG in second case is large enough for ``true'' MBS to emerge in this part.
For simplicity and without a loss of generality we take $\mu = 0$.
As previously, in both results, $2\Delta_\text{eff}$ and $G$ have a well visible parabolic-like boundaries between trivial and non-trivial phase.
However, in this case, we observed an additional horizontal line with small value of $2\Delta_\text{eff}$ with corresponding $G = 2 G_{0}$.
This results are independent of $V_\text{BG}$ and are associated with an existence of non--trivial topological phase at non-BG region of NW.
Moreover, its position can be modified by $\mu$.

From comparison of both results (cf.~Fig.~\ref{fig.res} and Fig.~\ref{fig.res2}), we can conclude that the length of the BG region plays similar role as the length of the homogeneous system. 
Indeed, in practice, the BG region can be treated as a ``new'' NW separated from ``old'' NW.
In our case, the non-homogeneous system can be represented as a few NW with different values of chemical potential connected together.

\section{Summary}
\label{sec.sum}

The Majorana bound states are characterized by the nearly--zero excitations, which values depends on the length of the nanowire (Fig.~\ref{fig.res3}). 
One of signatures of existence of the ``true'' Majorana bound states in the system can be observed  as the doubled quantum of conductance  $G = 2 G_{0}$.
This quantities can be useful in experimental reproduction of topological phase diagrams, where boundary between trivial and non--trivial topological phases are given by $h_{c} = \sqrt{ \tilde{\mu}^{2} + \Delta^{2} }$.

Typically, condition for the emergence of the topological phase depends on ``internal'' parameters of the system, i.e. on the chemical potential $\tilde{\mu}$ or superconducting gap $\Delta$.
In this study, we show a differential conductance $G$ in the case of the homogeneous system (Fig.~\ref{fig.res}).
Contemporary experimental techniques allow for local modification of $\tilde{\mu}$ by ``external'' parameter, i.e. gate potential $V_\text{BG}$.
Therefore, by tuning  $V_\text{BG}$ (and independently of $\tilde{\mu}$), the main properties of the topological phase diagrams can be reproduced (Fig.~\ref{fig.res2}).
Our calculation has been performed by modeling the experimentally available system, which was recently implemented ~\cite{chen.yu.17}  to confirm a realization of the topological phase diagram in the superconductor/semiconductor nanowire devices.
Both results, theoretical and experimental, confirm the emergence of the topological phase in such class of systems.

\begin{acknowledgments}
We thank Pascal Simon for inspiriting discussion.
This work was supported by the National Science Centre (NCN, Poland) under grants 
UMO-2018/31/N/ST3/01746 (A.K.), 
and
UMO-2017/25/B/ST3/02586 (A.P.). 
\end{acknowledgments}

\bibliography{biblio}

\end{document}